\documentclass[12pt,a4paper]{article}
\usepackage[utf8]{inputenc}
\usepackage[margin=1.5cm]{geometry}
\usepackage{amsfonts, amsmath, amssymb, amsthm}        
\usepackage{graphicx}
\usepackage{amsthm}
\usepackage{natbib}
\renewcommand{\cite}{\citep}

\title{Fractal scaling and the aesthetics of trees}
\author{Jingyi Gao$^{a}$ and Mitchell Newberry$^{b,c,d,*}$}
\date{}

\begin{document}

\maketitle

\begin{centering}\footnotesize
$^a$ University of Wisconsin, Madison, Department of Computer Science\\
$^b$ University of New Mexico, Department of Biology\\
$^c$ University of Michigan, Center for the Study of Complex Systems\\
$^d$ Max Planck Institute for Evolutionary Anthropology, Department of Human Behaviour, Ecology and Culture\\
$^*$ To whom correspondence should be addressed: \texttt{mgnew@unm.edu}\\
\end{centering}

\begin{abstract}
Trees in works of art have stirred emotions in viewers for millennia. Leonardo
da Vinci described geometric proportions in trees to provide both guidelines
for painting and insights into tree form and function. Da Vinci's Rule of trees
further implies fractal branching with a particular scaling exponent $\alpha =
2$ governing both proportions between the diameters of adjoining boughs and the
number of boughs of a given diameter. Contemporary biology increasingly
supports an analogous rule with $\alpha = 3$ known as Murray's Law.  Here we
relate trees in art to a theory of proportion inspired by both da Vinci and
modern tree physiology. We measure $\alpha$ in 16th century
Islamic architecture, Edo period Japanese painting and 20th century European
art, finding $\alpha$ in the range 1.5 to 2.5. We find that both conformity and
deviations from ideal branching create stylistic effect and accommodate
constraints on design and implementation. Finally, we analyze an abstract tree
by Piet Mondrian which forgoes explicit branching but accurately captures the
modern scaling exponent $\alpha = 3$, anticipating Murray's Law by 15 years.
This perspective extends classical mathematical, biological and artistic ways
to understand, recreate and appreciate the beauty of trees.
\end{abstract}

Classical theories of proportion, such as the rule of thirds or golden
sections, describe ratios between particular dimensions in a design. Vitruvius,
for example, prescribed specific proportions in architecture by relating them
to the human form, as iconified in da Vinci's Vitruvian Man.

\begin{figure}[t]
\centering
\includegraphics[width=180mm]{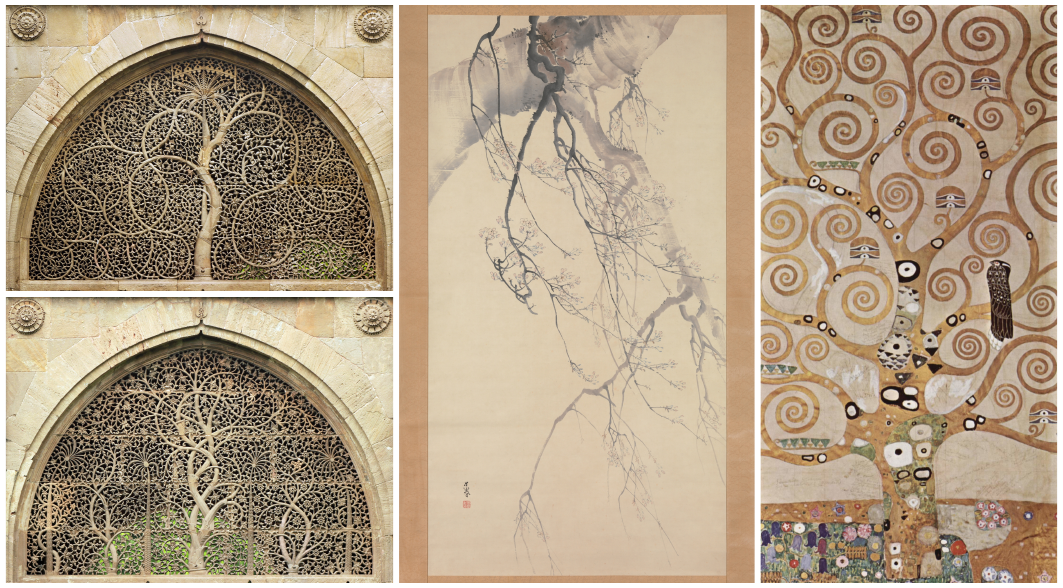}
\caption{Trees in artwork---whether realistic or highly stylized---exhibit
characteristic ratios between the sizes of branches that are both aesthetic and
representational.  Left: carved stone window screen of Sidi Saiyyed Mosque,
Ahmedabad, Gujarat, India (1573 CE). Center: \textit{Cherry Blossoms},
Matsumura Goshun (1752-1811), ink on paper.  Right: \textit{L'Arbre de Vie}
(Tree of life), Gustav Klimt (1909), oil on canvas.}
    \label{fig:works}
\end{figure}

Trees, by the nature of their variability, do not conform to prescribed ratios.
Each tree bends and branches as it grows, producing unique but recognizable
forms, with variation and detail that defy simple ratios. Yet trees in art,
such as the examples in Figure~\ref{fig:works}, make striking visual
impressions through careful attention to proportion.

Da Vinci also discusses proportions in trees, but rather than specify exact
proportions, he graphically describes a spectrum of possibilities. Da Vinci's
theory is prescient in its correspondence to modern tree physiology, which uses
a version of ``da Vinci's Rule'' with relatively minor modification
\cite{mcculloh2005patterns}. Here, we develop da Vinci's theory according to
its original intent---as a theory proportion in the arts---using contemporary
insights from tree physiology and fractal geometry that were unavailable in da
Vinci's time. We then compare the theory to famous works of art. We find works
that correspond to da Vinci's Rule, works that fall short due to design
constraints, and even works that insightfully anticipate the modern theory.

What distinguishes proportion in trees from proportion in the human form, Roman
columns or composition on a canvas, is branching. Each time a bough
branches into two, it creates smaller versions of itself, producing arbitrarily
fine structure and detail down to the finest twigs. Nor does this branching
follow deterministic rules; the branching pattern responds to subtle influences
of light, wind, temperature and gravity, creating unique ratios for every tree
in every situation as it grows.

Nonetheless, such recursive or self-similar forms are now understood as fractal
\cite{mandelbrot1977fractals}. Fractal geometry describes shapes with
ever-increasing detail at finer and finer scales using the idea of statistical
self-similarity: objects consisting of random, inexact smaller versions of
themselves, such as trees generated using recursive algorithms
\cite{prusinkiewicz2012algorithmic}. Fractal patterns have long appeared in art
and held aesthetic and spiritual value across cultures.  Traditional African
and Indian art, architecture, and cosmology are replete with fractal shapes,
self-similarity, and self-reference
\cite{eglash1999african,dutta2014symbolism}, while ancient Chinese calligraphy
critics prized script that ineffably resembled spectacular summer clouds, the
cracks in a wall and water stains of a leaking house---each prototypical
examples of fractals \cite{li2008fractal}.

Much effort has been devoted to measuring works of art according to the
definitive property of fractals---the fractal dimensionality
\cite{zeide1991method,taylor2007authenticating}---yet the results have been
sometimes controversial and not particularly informative
\cite{bountis2017fractal}. The focus on fractal dimensionality comes from
foundational discussions in mathematics on the relationships between dimension,
infinity and continuity \cite{hausdorff1918dimension,cantor1883ueber}, yet for
describing real systems, we posit that a more nuanced measure may be more
useful. Indeed, da Vinci's theory implies a scaling relationship between limb
radius and number which we describe below. This scaling relationship is akin to
fractal dimension, in that it governs how much detail appears with increasing
powers of observation, but it also corresponds to measurable physical
parameters in a modern theory of vascular biology
\cite{west1999fourth,zamir2005physics,newberry2015testing}.

Using da Vinci's approach, coupled with insights from fractal geometry and
statistics developed in biology \cite{newberry2019self}, we produce a theory of
proportion in trees that is measurable in art and comparable to results across
fields. Artistic idealizations of trees, in comparison to mathematical
idealizations \cite{mcculloh2003water} may be of interest even in biology,
where measurement is difficult due to broken branches, shade, bark, and other
complexities of the real world \cite{bentley2013empirical}. We therefore present an analysis in
terms of art, art history, biology and fractal geometry for the benefit of
each.

\begin{figure}[t]
\centering
\includegraphics[width=180mm]{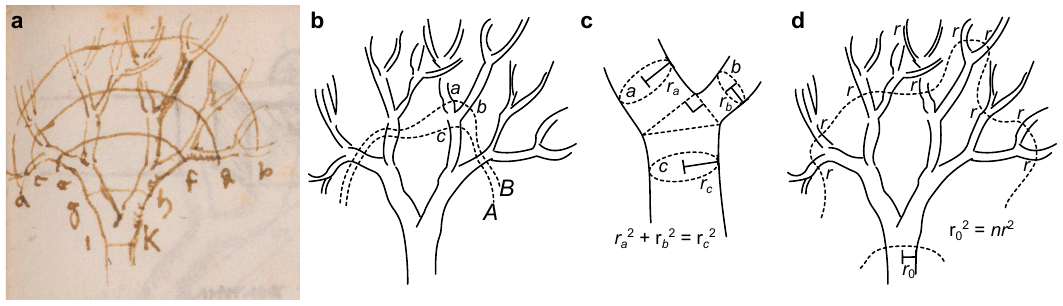}
\vspace{-1.5\baselineskip}
\caption{(a) da Vinci's sketch of a tree (from Institut de France Manuscript M,
p. 78v) shows that combined thickness is preserved at different stages of
ramification. Sketch (b) shows that if $A$ and $B$ have the same combined
thickness, $a$ and $b$ must have the same thickness as $c$. This implies that
(c) combined cross-sectional area ($\pi r^2$) is preserved across branching.
Likewise, (d) uses the principle to count the number of branches $n$ that have
radius $r$.}
\label{fig:davinci}
\end{figure}

\section{Theory of proportion from the physics of trees}

Our theory begins with a diagram in a da Vinci's account \textit{Botany for
Painters and Elements of Landscape Painting} \cite{richter1883literary}, which
we reproduce in Figure~\ref{fig:davinci}a. Da Vinci explains,
\begin{quotation}
``Every year when the boughs of a tree have
made an end of maturing their growth, they will have
made, when put together, a thickness equal to that of
the main stem; and at every stage of its ramification
you will find the thickness of the said main stem; as:
\textit{i k}, \textit{g h}, \textit{e f}, \textit{c d},
\textit{a b}, will always be equal to each other'' 
\end{quotation}
The diagram contains all the conceptual tools necessary to derive modern
principles of tree architecture using primarily visual reasoning. First, we can
show that thickness is preserved every time one bough branches into two. The
transects $A$ and $B$ in Figure~\ref{fig:davinci}b both represent cross sections that exactly
separate the leaves from the roots just as da Vinci's sections. $A$ and $B$
differ only in whether they incorporate $a$ and $b$ versus $c$, and so the
combined thickness of $a$ and $b$ must equal $c$. Scholars have interpreted
thickness as cross-sectional area as shown in Figure 2c. This area is $\pi
r^2$ for a branch whose radius is $r$, and so the equation
\begin{equation}
r_a^2 = r_b^2 + r_c^2
\label{eq:davinci}
\end{equation}
relates the radius $r_a$ of branch $a$ with $r_b$ of $b$ and $r_c$ of $c$, as
the factors of $\pi$ cancel from both sides. This equation is known in tree
physiology as da Vinci's Rule \cite{mcculloh2005patterns}.

Rather than prescribe a strict ratio to relate two dimensions, such as the
golden ratio $\phi$ might relate width and height, da Vinci's Rule allows
flexibility in how the thickness of all three branches relate. Some branches or
offshoots may be smaller or larger than others, but if one branch is smaller,
the other must be larger according to the formula. This constraint maintains
overall proportions of trees. Conveniently, the Pythagorean Theorem shares the
same formula, allowing an easy visualization of the possible relationships
between branch diameters: They are the same as the possible relationships
between the sides of a right triangle (cf. Figure~\ref{fig:crux}, middle row).

The diagram further allows us count how many branches have a certain approximate
radius $r$, which relates to fractal geometry. We draw a cross section that
cuts the tree only where the branches reach a size closest to $r$, as in Figure
2d. These $n$ cross-sections each with area $\pi r^2$ must sum to equal
the main stem's area $\pi r_0^2$, and so we have the approximate formulas
\begin{equation}
n r^2 \approx r_0^2 \quad\mbox{or}\quad n \approx \left({r_0\over r}\right)^2.
\label{eq:scaling}
\end{equation}
This describes a key property of a tree's fractal geometry: how much more
detail exists when we sharpen the clarity of observation \cite{mandelbrot1967long}. If we choose $r$ to
be the smallest observable radius in a photograph, the number of branches $n$
measures the amount of detail visible at this smallest scale. If we then
increase the camera resolution $x$-fold, then branches of size $r/x$ will now
be visible, revealing $x^2$ times more of them. Equation~\ref{eq:scaling} and
any equation of the form $y = cx^\alpha$ is called a scaling relationship, and
fractal dimension is also a scaling relationship \cite{mandelbrot1977fractals}.
Here the scaling is quite rapid: a two-fold increase in resolution reveals four
times more branches.

One wrinkle remains in making da Vinci's explanation consistent with modern
science. In da Vinci's statement, the combined thickness is preserved across
ramifications. What is truly preserved is water. Fluid entering one section of
branch must equal the volume of fluid leaving the other side, which also
enters the downstream branches. Thus it is combined fluid flow, rather than the combined thickness, that is
equal at every stage of ramification. Whether flow corresponds to cross-sectional
area, however, is another question. Da Vinci gives good reason to equate the two, later writing ``all the branches of a watercourse at every stage of its
course, if they are of equal rapidity, are equal to the body of the main
stream." The wrinkle, however, is that we no longer believe that water in
trunks and twigs moves with equal rapidity. Water tends to slow somewhat as it
approaches the leaves from the trunk. This is possible if the combined
thickness increases, because a greater volume of slower-moving water has the
same total flow. The net result is that fluid flow is not necessarily
proportional to thickness or $r^2$. Indeed, Cecil Murray in the early 20th
century reasoned that biological evolution should optimize fluid flow to be
proportional to $r^3$ rather than $r^2$ \cite{murray1926vascular}. Modern
tree physiologists find that the exponent 3 may be closer to the truth
\cite{mcculloh2003water}. Measurements vary from tree to tree, however, and
within the same tree over its lifespan \cite{bentley2013empirical}. Fluid flow
within trees is still an active research topic \cite{mcdowell2019hydraulics}.

Fortunately, our earlier arguments and da Vinci's picture still hold if we
replace $r^2$ with $r^\alpha$. We simply use a formula for flow $f = cr^\alpha$
instead of $a = \pi r^2$ and carry the argument through as before. In deference to the mysteries of trees, we leave the
constant $c$ and the exponent $\alpha$ as unknowns\footnote{the constant $c$ immediately disappears just as 
$\pi$ disappears from Eq.~\ref{eq:davinci} and
thus has no influence on the result.}. The result is a generalized version of the
equations, \begin{subequations}
\begin{align}
r_a^\alpha =& r_b^\alpha + r_c^\alpha \quad\mbox{ and} \label{eq:generala}\\
n =& \left({r_0\over r}\right)^\alpha \quad\mbox{.}
\label{eq:generalb}
\end{align}
\end{subequations}
Here we assume only that fluid flow is proportional to some unknown power of
radius. Setting $\alpha = 2$ gives da Vinci's Rule, while $\alpha = 3$ gives
Murray's Law, but we recognize that these are idealizations. The truest
$\alpha$ for any given tree might be more or less or somewhere in between.
Furthermore, the conserved power of radius in Equations \ref{eq:generala} and
the scaling relationship Equation \ref{eq:generalb} have one and the same
exponent. Thus, $\alpha$ specifies both the proportion of limbs as they branch
and the scaling relationship between radius and the number of branches, even
considering randomness and asymmetry in the development of a tree.

\begin{figure}[t]
\centering
\includegraphics[width=90mm]{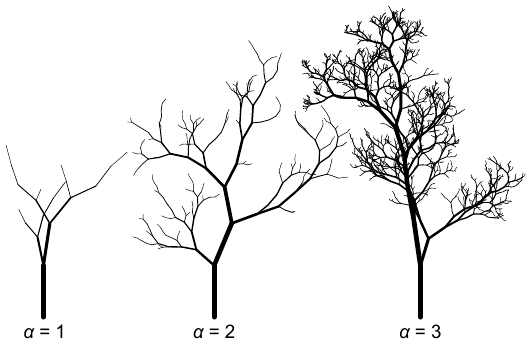}
\caption{The scaling exponent $\alpha$ determines the amount of detail
present for a given contrast in size between the smallest and largest branches.
Each tree above is generated by the same random algorithm starting from an identical
trunk, differing only in the value of $\alpha$ used to compute the branch
diameters by solving Equation \ref{eq:generala}.  The algorithm stops adding
branches when the radius reaches roughly 1/10th that of the trunk. Taking
$r_0/r = 10$ in Equation \ref{eq:generalb} gives 10 branches for $\alpha = 1$,
100 branches for $\alpha = 2$ and 1,000 for $\alpha = 3$. The trees above vary
randomly around these numbers, since no branch radius is exactly $r$.
Scientists believe the $\alpha$ of natural trees falls in an interval including
2 and 3.}
\label{fig:grove}
\end{figure}

\section{Aesthetics of $\alpha$}

The most striking consequence of $\alpha$ is its role in fractal scaling,
controlling the amount of detail for a given contrast between the largest and
smallest branch sizes. Figure~\ref{fig:grove} depicts three trees, each with a
10 fold difference in diameter between the largest and smallest branches.
Because $\alpha=1$ leads to a rapid reduction in branch thickness according to
Equation \ref{eq:generala}, relatively few branches exist in this size interval. Taking $\alpha = 3$ produces more gradual reduction in branch size, and
hence more ramifications. Each ramification in turn leads to twice as many
downstream branches, and hence hence increasing $\alpha$ gives exponentially
more detail.

\begin{figure}[t]
\centering
\includegraphics[width=90mm]{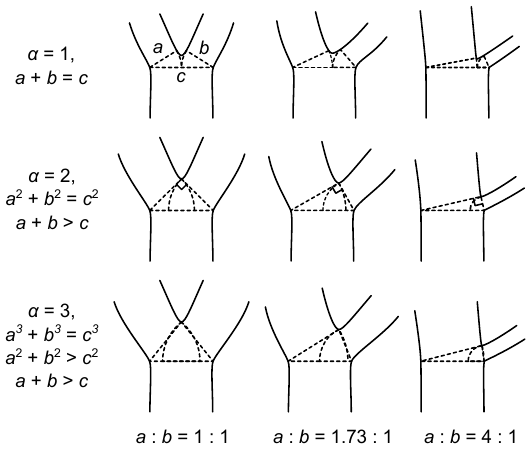}
\caption{$\alpha$ constrains the proportions among adjacent branches according
to Equation~\ref{eq:generala}. Here the diameters of the downstream boughs are
labeled $a$ and $b$ and the upstream bough is $c$. Each row depicts a different
$\alpha$ and each column a different ratio between $a$ and $b$.
Equation~\ref{eq:generala} does not specify exact proportions between each
bough, but merely a spectrum of possible relationships. If we specify the
proportion $a : b$, however, then Equation \ref{eq:generala} provides unique
proportions between all three branches. The diameter $c$ is the same in each
picture, but $a$ and $b$ (the left and right branches respectively) differ
depending on $\alpha$ and the proportion $a : b$.}
\label{fig:crux}
\end{figure}

The exponent $\alpha$ also prescribes mathematical proportions among individual
boughs as they branch, as in Figure \ref{fig:davinci}c and
Equation~\ref{eq:generala}. Here $\alpha$ determines the power of radius or
diameter that is preserved across branching. $\alpha = 1$ means that diameter
itself is preserved, so that the sum of diameter of the two downstream branches
is equal to the upstream branch as shown in the top row of
Figure~\ref{fig:crux}. $\alpha = 2$ means that the square of diameter, which is
proportional to the cross-sectional area, is preserved, with the consequence
that the combined diameter increases. The relationships between branch diameters
for $\alpha = 2$ are easy to visualize, since they exactly correspond to the
possible relationships between sides of a right triangle, such as the three
triangles in Figure~\ref{fig:crux}'s middle row. If $\alpha = 3$, the total
area and diameter both increase across branching, corresponding to slowing flow
of water within the tree.

Yet here the visual influence of $\alpha$ is more subtle. The proportions in
the first two rows of Figure~\ref{fig:crux} might appear identical to casual
observation, but differ enough to a painterly eye for da Vinci to remark on the
constraint and argue for $\alpha = 2$. The differences between the second two
rows, however, are hardly discernible without precise measurements or
mathematical analysis.  Geometric constructions for cubic ratios, involved in
the triangles in the third row, are famously impossible\footnote{The problem of
``doubling the cube", identical to solving for the ratios in the bottom left
picture, was studied since antiquity, and later proven to be impossible to
solve with classical straitedge and compass geometry \cite{lutzen2010algebra}.
Fortunately, solutions can be easily computed with modern numerical methods and
algebra.}, nor was the scaling of fractal branching elaborated until the 20th
century, leaving da Vinci with no obvious method to discern the exponent 2 from
3. Nonetheless, these subtle differences compound exponentially over
generations of branching to create striking differences in the amount of detail
in a rendering, as seen in Figure~\ref{fig:grove}.

\section{Scaling of trees in art}

We measured $\alpha$ in trees in famous artworks, as scientists have done for
natural trees. We chose the three artworks in Figure~\ref{fig:works} as
demonstration cases, for their variation in degree of realism, culture of
origin, time period and medium. We chose works that depicted trunks or large
branches offering sufficient contrast between the largest and smallest branches
to measure scaling. We also concede, despite statistical precepts to the
contrary, that we chose these works because they made us feel awe for the
beauty of trees. Fortuitously, the works are all of comparable size, with the
largest boughs between 10 and 26 cm.

For each work, the two authors independently measured the diameter of each
branch according to rules we describe in Detailed Methods. We thereby produced
two lists of all branch diameters in each work to control for subjectivity in
assessing diameter and the presence or absence of boughs and branching. We have
found in previous work \cite{lin2023seeing} that counting the number of branches at
different magnitudes of diameter is a far more accurate measure of $\alpha$
than measuring proportions at individual branch points
\cite{newberry2015testing}. This fact is also evident in the contrast between
Figures~\ref{fig:grove} and \ref{fig:crux}. Therefore we do not even record
which diameters correspond to which branches, saving tremendous labor.

\begin{figure}[t]
\setkeys{Gin}{width=180mm}
\includegraphics[width=180mm]{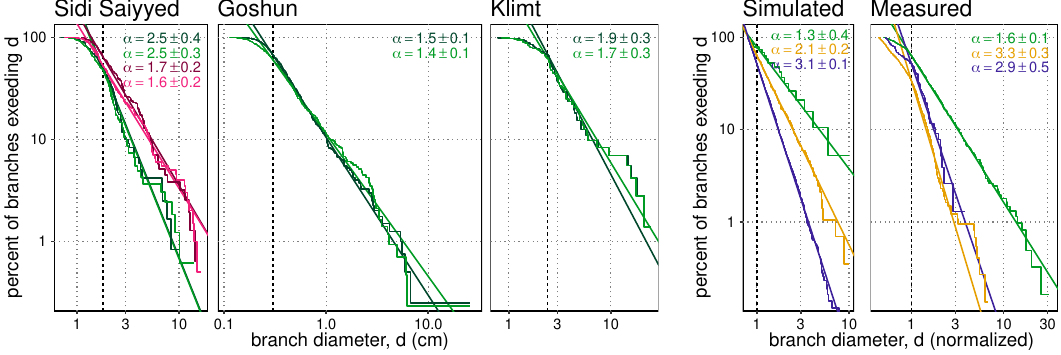}
\caption{Data and inference of $\alpha$ for the works in Figure~\ref{fig:works}
along with simulated and measured trees. Each plotted line shows, for every
value $d$ on the horizontal axis, the percent of branches that are greater than
$d$, ranging from 100\% to (100/$N$)\% where $N$ is the total number of
branches. The slope of each line is equal to $-\alpha$. Curvature indicates
departures from self-similarity, while jagged steps indicate few measurements,
statistical error or measurement error. Each author scored the tree(s) in each
work, producing two replicate lines for each tree. Sidi Saiyyed Mosque features
two large trees, which we plot separately. Simulated diameters come from the
trees in Figure~\ref{fig:grove}, each with a different value of $\alpha$.
Measured diameters come from public data on balsa, pi\~non and ponderosa trees
that were sacrificed and measured for science \cite{bentley2013empirical}. Dotted
vertical lines indicate a minimum threshold for data to be included in the
statistical inference. Simulated and measured datasets are divided by this
minimum value and therefore the horizontal axis has no units.
}
\label{fig:results}
\end{figure}

We plot each list of measurements in Figure~\ref{fig:results}, with diameter on
the horizontal axis and the percentage of diameters that exceed that diameter
on the vertical. That is, as diameter $d$ increases on the horizontal axis, a
smaller and smaller percentage of the branches exceed $d$, and so the curve
descends until finally no branches exceed $d$ and the curve ends. The reason
for this specific plotting method is that the overall slope of the line
then corresponds mathematically to $-\alpha$ when both axes are plotted using
logarithmic scales \cite{newberry2019self}. This allows us to read $\alpha$ off
the plot as well as measure it statistically. The plots therefore contain all
diameters in each data set, and we print the corresponding statistical
estimates $\hat\alpha$ with 95\% confidence intervals. These plots reveal
self-similarity as straight lines (with slope equal to the scaling exponent),
deviations from self-similarity as persistent curvature in one direction, and
random variation and measurement error as wiggles or jaggedness. Each plot
then provides a
compact but comprehensive picture of how well the generalized da Vinci model
fits the branch diameter measurements. For comparison, we also show the radii
of the simulated trees from Figure~\ref{fig:grove} and three more natural trees
that were sacrificed and measured for science \cite{bentley2013empirical},
illustrating conformity to Murray's law as well as the scope of individual
variation \cite{mcculloh2004murray}.

Medieval Islamic design and architecture is known for precise and sophisticated
mathematical patterns \cite{lu2007decagonal}. The stone window screen (jali) of
the late medieval Sidi Saiyyed Mosque pictured in Figure~\ref{fig:works} top
has a scaling exponent of 2.5$\pm$0.4, midway between da Vinci's analysis and
the contemporary scientific expectation, with branch diameters that conform to
the fractal branching model comparably well to real trees. The second jali
pattern consists of five trees. We plot the diameters in
Figure~\ref{fig:results} both excluding and including the smaller trees. In
either case, the statistical estimates fall within the range 1.6-1.7$\pm$0.2.
These estimates differ substantially from the first jali, by nearly a whole
number. The second jali constrains its main tree to 50\% of the total area,
though its stem begins slightly larger than the first jali at 16 versus 14 cm.
This alternative design leaves no room for the dramatic spirals of the first
jali, with the result that branching is concentrated in the larger branches,
distorting true self-similarity. If we instead draw a straight line from the
largest to smallest branches in Figure~\ref{fig:results}, the slopes fall in
the range 2.1 to 2.2, marginally consistent with the first jali and da Vinci's
prescription. In both cases, the regularity of proportional decrease in the
branches shows conformity to da Vinci's theory, even as the exponent varies to
accommodate design constraints.

\textit{Cherry Blossoms} is an ink on paper painting by Matsumura Goshun
(1752-1811) of Edo period Japan. The diameters of Goshun's tree conform
exquisitely to self-similar proportions---with less departure even than real
trees---particularly among the most intricate branches between 3-10 mm. Though
extremely rich in detail, with over 400
individual branches, we estimate the scaling exponent for this painting at only
1.4-1.5$\pm$0.1, clearly excluding da Vinci's $\alpha = 2$. This combination of
extreme detail and rapidly decreasing branch diameter is made possible by the
extreme 200-fold contrast in diameter between the largest and smallest
branches. This extreme contrast in size highlights the delicacy of the smallest branches,
while the low value of $\alpha$ lends an austerity or emptiness to the
composition, enhanced by faintly rendered branches as if seen
distantly through fog. This $\alpha$ then represents a tradeoff between
dramatically exaggerated proportions, austerity and perhaps practicality:
Achieving $\alpha = 2$ across such large and small branches would require over
ten times as many branches, which would be labor intensive and might clutter
the figure.

Gustav Klimt's \textit{L'Arbre de Vie} (Tree of life) represents only a 30-fold
size difference between the smallest and largest branches---the least we study---and a highly stylized representation. While the scaling exponent estimates
are close to 2 (1.7-1.9$\pm$0.3), the trunk and the first and second branch are
larger than self-similarity would prescribe, leading to modest underestimates of
$\alpha$. \textit{L'Arbre de Vie} is one of seven equally sized panels in the
Stoclet Frieze, each containing branching spirals. Yet only this center panel
contains self-similar branching, while other others more simply repeat a spiral
motif with no further decrease in size. We measure only the central panel,
before the design fades into a repeating pattern more characteristic of
wallpaper tiling than fractals. We nonetheless find self-similarity among the branches between 2.4 and 5 cm, roughly half of the data.

\section{The treeless tree}

\begin{figure}[t]
\centering
\includegraphics[width=180mm]{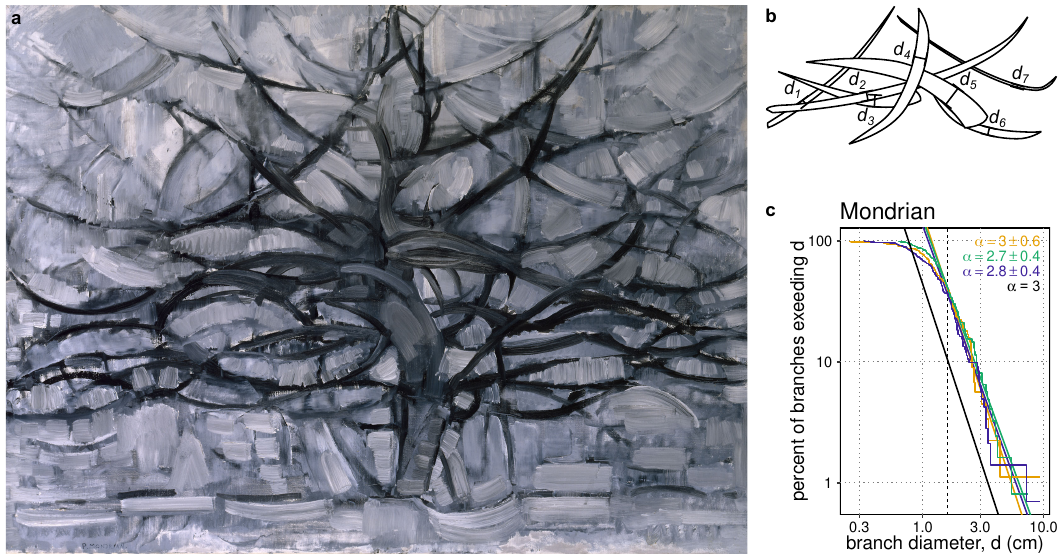}
\caption{(a) \textit{De grijze boom} (Gray tree) Piet Mondrian, 1911. (b)
Absent discernible branch points, we mentally decompose the figure into arcs
and record the approximate diameter of each arc near its visual center of
gravity. (c) the diameters and statistical estimates as in Figure~\ref{fig:results} for \textit{De
grijze boom}.}
\label{fig:mondrian}
\end{figure}

Branching is so essential to trees that the word tree has come to refer to
almost anything that branches, even metaphorically, such as a family tree or
decision tree. Diameters around branch points are indeed essential to
the development of our theory. Yet Piet Mondrian, pioneer and theorist of
20th-century abstract art, shows us a tree without branching. Mondrian's career
reveals a fascination with trees. Particularly from 1905 to 1912, he produced
many renderings of intricately branching trees, including paintings that have
been studied as
fractals \cite{bountis2017fractal}. In his 1911 \textit{De grijze boom} (Gray
tree, Figure~\ref{fig:mondrian}a), however, we see an abstract
cluster of dark arcs superimposed at varying angles against a gray
visual field, without obvious branching. The arcs intersect seemingly at
random without regard for proportion or identifiable upstream or downstream
connections, yet from this disconnected collection of brushstrokes emerges an
unmistakable image of a winter tree stark against a bleak sky.

How does Mondrian communicate the image of a tree so vividly without using 
the tree's most essential feature? We argue that \textit{De grijze boom} instead
captures a feature just as innately recognizable: a tree's essential fractal
scaling. To test this conjecture, we measure the diameters of Mondrian's
brushstrokes near the center of each arc (Figure~\ref{fig:mondrian}b), despite
that these ``limbs'' have no branching or anatomically relevant points of
intersection. Indeed, Mondrian's tree is the most faithful in representing the
contemporary scientific consensus on tree anatomy, achieving statistical
consistency with $\alpha = 3$. Mondrian's tree contains less detail than
most we study, but it is also less
precisely rendered. With a 30-fold range of limb size---comparable to Klimt's
\textit{Tree of Life}---Mondrian captures almost perfect self-similarity and
the relationships between precision and detail implied in
Figure~\ref{fig:davinci}d and Equation~\ref{eq:generalb}---among small and
large branches alike. Mondrian's exponent then concords with what physiologist
Cecil Murray subsequently hypothesized in 1926 \cite{murray1926vascular}, which
could only recently be measured accurately \cite{mcculloh2004murray}. Together
with curved lines and the impression of a horizon and a central trunk, we
suggest that the insightful and realistic scaling of diameter is integral to
what makes Mondrian's abstract strokes a tree.

\section{Discussion and conclusion}

We present new extensions to a classical perspective on how to appreciate and
recreate the beauty of trees. This perspective allows us to compare trees in
design with science, as well as provide, as da Vinci did, a theory of
proportion.  Our findings show how art often prefigures scientific
understanding and the two provide complementary lenses on the natural and human
worlds. Da Vinci's observations and the near-contemporaneous architecture of
Sidi Saiyyed show consilient understanding of branching patterns, while
Mondrian and Murray outline a shift in both scientific and artistic portrayal of
trees in the early 20th century.

There is no doubt that fractal shapes have been prized aesthetically, yet
relating fractal dimensionality \textit{per se} to particular aesthetic
qualities has been challenged. On the one hand, psychological studies have
found fractal shapes pleasing or reassuring
\cite{joye2007architectural,brachmann2017computational}, their appearance in
hallucinations suggests that certain fractals are innate to human visual
processing \cite{bressloff2001geometric}, and different dimensionalities are
recognizably different \cite{spehar2003universal}. Yet on the other, both
fractal and non-fractal objects have aesthetic merit and both pleasing and
displeasing forms exist at every fractal dimensionality
\cite{jonessmith2009drip}. This contradiction may arise because dimensionality
is altogether too coarse a measure. If painting is 2-dimensional and sculpture
is 3-, we do not argue one or the other superior, so why should we debate the
merits of having fractal dimensionality 1.6 versus 1.4? Our work rather
suggests that more specific measures in context are more likely to support
clearer relationships between fractal properties and physics, aesthetics or
perception.  Here we study radius scaling in trees, but we believe different
principles and measures would apply to feathers or snow flakes. Many relevant
properties have been measured even within trees and vascular networks
\cite{tekin2016vascular}, and these too should be employed in the arts and
sciences.

A curious classical exception to the theory is coppices and pollards. Coppicing
and pollarding involve pruning a tree at its base, trunk or large branches so
that the tree produces a recurring supply of new growth, such as for kindling,
fodder or wattle and daub construction. This practice was once commonplace in
Europe and continues throughout the world. Indeed, in the explanation we study
da Vinci continues ``unless the tree is pollard---if so the rule does not hold
good'' \cite{richter1883literary}. Pollarding disrupts a tree's natural growth
pattern and introduces a rift in its self-similarity.  Branch pruning leads to
a separation of scales: one large trunk and many small branches with nothing in
between. Yet pollards themselves have an interesting art history and dominate
compositions such as Mondrian's \textit{Willow grove: impression of light and
shadow} or the Song dynasty classic \textit{Along the River During the Qingming
Festival}. These exceptions may be as aesthetically interesting as the rule,
and may have once evoked romantic images of pastoral life or nostalgia in
audiences.

Although we study a diverse sample of artworks, our choice is neither expansive
nor systematic. We expect these methods to yield more insights into art
history when applied to a larger scope of work. For example, paintings of
lightning, akin to trees, have gotten more realistic over time
\cite{stromp2018realistic} and computational analyses of composition across
thousands of landscape paintings reveal proportional rules as well as
systematic evolution over time periods, styles and individuals
\cite{lee2020dissecting}.

We conclude that da Vinci's theory of trees---prescient, insightful and correct
up to the precision available in his time---can be fruitfully extended to
account for radius scaling, with attention to the scaling exponent. We thereby
produce a more modern interpretation of the same essential theory of proportion
informed by advances in fractal geometry and tree physiology. The theory
applies to painting as much as algorithmically generated trees in modern
graphics, and can be used to produces vivid, beautiful and realistic designs.
In comparing famous works to the theory, we find both interesting regularities
and exceptions. This perspective offers new ways to think about trees, and new
design principles to follow or break as art wills.

\section{Detailed Methods}

\textbf{Reproducible research.} All code and research materials for automated
reproduction of the analysis are available as a public git
repository\footnote{https://github.com/mnewberry/treescale}, including
original hand-annotated image files.

\textbf{Measuring branch diameter.} We annotate images of each work using lines
in the open source scalable vector graphics (SVG) editing program Inkscape. We
overlay line segments by hand perpendicular to the direction of the branch to
represent each branch diameter. We attempt to choose each diameter as the
closest diameter downstream of each branch point that represents the overall
branch diameter. That is, we measure diameter downstream of any transient
changes in diameter such as the concave curves at Klimt's branch points or the
leaves in Sidi Saiyyed. We then load the SVG file in Mozilla Firefox use
JavaScript
code\footnote{\texttt{Array.prototype.slice.call(document.getElementsByTagName("path")).map(function(a)
\{ return a.getTotalLength() \}).join("\textbackslash{}n");}} in the Web
Developer Console to extract the branch points, which we paste into a file. We
selected publicly-available images that clearly showed the works, avoiding
avoiding parallax error or shadows that might obscure the branch thickness. We
stopped annotating branches when small stems lead only to a single leaf or
motif. That is, we do not count as branches the leaves or flowers in Sidi
Saiyyed or Goshun or the Egyptian revival decorative motifs in Klimt.

In Mondrian's tree, anatomical branch points are indistinct or nonexistent.
Instead, boughs are represented by long, curved brush strokes without
anatomically relevant points of intersection. Therefore rather than apply the
scoring rules at branch points as with the other works, we measure the
diameters of each arc without regard to how the arcs intersect. We mentally
decompose the tree into arcs (as in Figure~\ref{fig:mondrian}b) and attempt to
measure the diameter near the center of each arc. We interpret each continuous,
regular dark curve as an arc, whether it is a single dark brush stroke, an
absence of light brush strokes, or a discernible dark shadow underneath gray
brush strokes. We roughly require each arc to curve in the same direction, so
that we decompose Y- or S-shaped patterns as two or more arcs on top of each
other. For consistency, we try to measure the arc near its visual ``center of
gravity'', such as its midpoint, thickest point, or somewhere in between. As
the process is somewhat subjective, we replicate the scoring process across
three independent viewers given only these instructions.

Each author independently scored each image, resulting in two replicates that
show the extent of researcher subjectivity in interpreting the images and
scoring rules. Our data files and annotated SVG files are available in the
git repository. For Mondrian's painting, we solicited a third ``blinded''
replicate from an anonymous participant 'a' who was given only an excerpt from
this methods section and Figure~\ref{fig:mondrian}a-b.

\textbf{Statistical estimation.} We fit pooled branch diameters to a discrete
power-law distribution using a maximum-likelihood method appropriate for
branching data \cite{lin2023seeing} following all relevant guidelines in
selecting the parameters $\lambda$ and $x_m$. This method involves binning the
data according to powers of $\lambda$ to reduce the influence of within-scale
noise.  We use $\lambda = 2$. As necessary when fitting power laws
\cite{clauset2009power}, we ignore data below a certain threshold, often called
$x_m$ in the literature on power law inference. Data below this threshold are
two small to be reliable either due to measurement error or random censorship
in the underlying process, such as an artist's arbitrariness in deciding
whether a branch is too small to paint. We chose these thresholds for each work
by examining the works themselves and deviations from self-similarity in the
data: 1.8 cm for Sidi Saiyyed, 3 mm for Goshun, 2.4 cm for
Klimt, and 1.6 cm for Mondrian.  The minimum criterion based on the work was
that the diameters are reliably measurable. The minimum criterion from the data
was that the threshold excluded curvature at low values in the plots in
Figure~\ref{fig:results}, as this curvature violates the self-similar model and
biases the estimation procedure.  A quick visual check that the threshold is
sufficient is that the slope of the curve matches the slope of the inference at
the threshold. We verified that this is true for all inferences. Measurements
from real trees frequently violate self-similarity, particularly for small
branches as bark and other anatomical specifics begin to contaminate the
measurements.  Hence we chose a separate threshold for each individual tree
that produced good fits to the largest branches. In each case this was also the
minimum threshold that passed the visual check.

\end{document}